# A rhombohedral ferroelectric phase in epitaxially-strained $Hf_{0.5}Zr_{0.5}O_2$ thin films

Yingfen Wei[1], Pavan Nukala[2], Mart Salverda[1], Sylvia Matzen[3], Hong Jian Zhao[4], Jamo Momand[1], Arnoud Everhardt[1], Graeme R. Blake[1], Philippe Lecoeur[3], Bart J. Kooi[1], Jorge Íñiguez[4], Brahim Dkhil[2], Beatriz Noheda[1,*]

[1]Zernike Institute for Advanced Materials, University of Groningen, 9747 AG Groningen, The Netherlands

[2]Laboratoire Structures, Propriétés et Modélisation des Solides, CentraleSupélec, CNRS-UMR8580, Université Paris-Saclay, 92295 Châtenay-Malabry, France.

[3]Center for Nanoscience and Nanotechnology, CNRS-UMR 9001, Université Paris-Saclay, 91405 Orsay, France

[4] Materials Research and Technology Department, Luxembourg Institute of Science and Technology (LIST), 5 avenue des Hauts-Fourneaux, L-4362 Esch/Alzette, Luxembourg

* E-mail: b.noheda@rug.nl;

**After decades of searching for robust nanoscale ferroelectricity that could enable integration into the next generation memory and logic devices, hafnia-based thin films have appeared as the ultimate candidate because their ferroelectric (FE) polarization becomes more robust as the size is reduced. This exposes a new kind of ferroelectricity, whose mechanism still needs to be understood. Towards this end, thin films with increased crystal quality are needed. We report the epitaxial growth of $Hf_{0.5}Zr_{0.5}O_2$ (HZO) thin films on (001)-oriented $La_{0.7}Sr_{0.3}MnO_3$/$SrTiO_3$ (STO) substrates. The films, which are under epitaxial compressive strain and are predominantly (111)-oriented, display large FE polarization values up to 34 $\mu C/cm^2$ and do not need wake-up cycling. Structural characterization reveals a rhombohedral phase, different from the commonly reported polar orthorhombic phase. This unexpected finding allows us to propose a compelling model for the formation of the FE phase. In addition, these results point towards nanoparticles of simple oxides as a vastly unexplored class of nanoscale ferroelectrics.**

Ferroelectric (FE) materials, exhibit switchable spontaneous polarization, and are of great technological interest for a myriad of applications, notably as memories and field-effect transistors in microelectronics[1], spintronics[2], and micro/nano electro-mechanical systems[3]. However, miniaturizing ferroelectrics is not an easy task, since depolarization fields become increasingly important at reduced sizes.[4,5] Searching for robust FE properties at the nanoscale has been, therefore, a recurrent challenge over the past decades. Thus, the recent discovery of ferroelectricity in ultrathin layers of the $HfO_2$-based materials[6] represents a real breakthrough in the field, and seems to expose a new type of ferroelectricity: one that only appears at the nanoscale and becomes better at smaller dimensions. In addition, silicon compatibility, the simplicity of their chemistry and low toxicity make them very attractive compared to the other commonly used FE layers.

In bulk, the stable form of $HfO_2$ (and $ZrO_2$) based compounds is a monoclinic phase ($P2_1/c$, m-phase) at room temperature.[7,8] Other common high-temperature and high-pressure phases, namely, tetragonal ($P4_2/nmc$, t-phase) and cubic (Fm-3m, c-phase) phases,[7,8] can be stabilized at room temperature via doping or nano-structuring.[9–11] In addition, rhombohedral phases (r-phase) have also been obtained by doping and applying mechanical stress.[12–15] The t- and r-phases are distortions from the fluorite structure (c-phase) and differ from the m-phase in their significantly smaller volume and higher cation coordination (8 instead of 7). None of the abovementioned phases are reported to be polar.

A polar orthorhombic phase ($Pca2_1$, o-phase) was first reported for Mg-doped $ZrO_2$ when cooling to cryogenic temperatures.[16] This polar phase is now believed to be the structural origin for the recently reported ferroelectricity in $HfO_2$-based thin films.[6] Recent literature has gathered examples of these FE films with different dopants (i.e Al[17], La[18], Zr[19], Si[20,21], Gd[22], Sr[23], Y[24]), on different substrates (i.e Si[6], Y-$ZrO_2$[24]), with different electrodes (i.e TiN[25], Pt[26],

Ir[27], TaN[28]), and by different growth methods, such as atomic layer deposition (ALD)[25], chemical solution deposition (CSD)[29], pulsed laser deposition (PLD)[24], and chemical vapor deposition (CVD)[30]. Various possible mechanisms, such as stress[25,31], doping[22], confinement by the top electrode[32], interface diffusion[19] or surface energy[33,34], have been put forward as stabilizing factors for the FE phase.

Initially, the best FE properties were only reported in ultra-thin films (usually around 10 nm thick).[20,35,36] The polarization declined significantly with the non-polar monoclinic phase appearing, when the film thickness increases.[36] Later on, FE films with a thickness of 50 nm have been achieved by laminating the HZO (each layer being 5 nm thick) with interlayer $Al_2O_3$.[37] Ferroelectricity has also been found in very thick films (390 nm), where the average grain size is below 10 nm.[29] Thus, all reported FE hafnia-based films have in common that they are formed by small crystallites, which shows clear evidence of the crucial role played by size effects in stabilizing the FE phase.[19,38] Indeed, it is known that in nanoparticles of radius $r$, the surface energy ($\sigma$) can produce large internal pressures ($P = 2\sigma/r$) of the order of GPa.[39,40] Thus, small crystal grains will prefer the room temperature stability of lower volume c- or t-phases than the m-phase.[38,41,42] For thicker films, usually above 10 nm (where the crystals have the possibility to grow further), the monoclinic bulk phase is always present. The polar o-phase has been postulated as the transformation phase between the t-phase and the m-phase.[16,28,43]

Most works report on ALD-grown films, which are polycrystalline and contain multiple phases (m-, t-, o-phases). In addition, the similarity of these structures and the small size of crystallites make a complete structural characterization even more challenging. Therefore, well-oriented samples, preferably in a single-phase are desired to study the factors responsible for ferroelectricity. Single-crystal, epitaxial Y-doped $HfO_2$ films have been achieved by PLD on yttrium oxide-stabilized zirconium oxide (YSZ) substrates with the polar o-phase,[24] reaching a polarization of 16 μC/cm². Here, we also utilize PLD to grow highly-oriented FE HZO films on a perovskite $SrTiO_3$ (STO) substrate with $La_{0.7}Sr_{0.3}MnO_3$ (LSMO) as a back-electrode, in order to gain insights into the precise role of strain (imposed by the substrate) on the ferroelectricity in hafnia-based systems. A polar rhombohedral phase has been discovered as the origin of ferroelectricity. In addition, we show that this phase solved one of the issues around the utilization of this material in devices: the need of "wake-up" pre-cycling.

X-ray diffraction (XRD) *θ-2θ* patterns along the crystal truncation rod of HZO films with different thicknesses are shown in Fig. 1a. Next to the (001) specular Bragg reflection of the STO substrate and the epitaxially-grown LSMO bottom electrode with thickness of ~30 nm, the main Bragg peak of the HZO films appears at around 30°. The crystal truncation rods are visible in the form of thickness oscillations, which demonstrates the good crystalline quality and interfaces of the films. This *2θ* value is slightly lower than that corresponding to the (111) reflection in the commonly reported polar o-phase in HZO, which appears at around 30.5° as shown by the black dashed line.[36] This indicates an expanded (111)-spacing ($d_{111}$) in the out-of-plane direction. In particular, extremely thin films below 4 nm (Fig. 1b) display highly elongated unit cells with the (111) reflection appearing at *2θ* well below 30°. As the thickness decreases, the HZO (111) peaks shift rapidly to smaller angles (larger d-spacing), which indicates a huge compressive in-plane strain for the thinnest layers. For films thicker than 9 nm, new peaks appear (at 28.3° and 34°), which can be assigned to the m- phase, consistent with the stabilization of the m- phase with increasing crystal size (while increasing thickness).[36]

According to XRD, the films are (111)-oriented (Fig. 1). Pole figure (texture) measurements were performed around the {111} peaks in a 9 nm-thick film. In a (111)-oriented single domain film, the other three reflections, (-111), (11-1) and (1-11) are expected at an (χ) angle of ~71° from the out-of-plane direction, with azimuthal angles (φ) differing by 120°. As seen in Fig. 2a, 12 reflections instead of 3 are found at $\chi \sim 71°$, revealing four crystallographic domains with different but well-defined in-plane orientations. The crystal domains are rotated 90° with respect to each other, following the four-fold symmetry of the (001)-oriented cubic substrate. In addition, four weak reflections at $\chi \sim 55°$ reveal a small amount of (001)-oriented component in the film. Synchrotron x-ray diffraction was used to scan all 13 reflections in the pole figure, revealing that the 12 peaks with in-plane components of the scattering vector (at $\chi \sim 71^0$) share the exact same $2\theta = 27.13°$ (Fig. 2b), giving rise to a d-spacing ($d_{11-1}=d_{1-11}=d_{-111}=2.94$ Å) that is significantly smaller than that of the out-of-plane (111) reflection ($d_{111}= 2.98$ Å). Thus, these measurements reveal a multiplicity that is only consistent with a rhombohedral unit cell (Fig. 2c) and a polar (three-fold) axis out of the plane of the sample.

To gain further understanding of the structure, we performed transmission electron microscopy (TEM) and local spectroscopy studies. Plane-view selected area electron diffraction (SAED), shown in Fig. 3a for a 9 nm thick film, displays a superposition of diffraction patterns from at least two domains of HZO. {220} spots from both domains occur at $\varphi=45°+60n$ (yellow circles in Fig. 3a) and at $\varphi=15°+60n$ (blue circles), with n being an integer from 0 to 5 (the $\varphi=0°$ direction is defined as the $[100]_{STO}$ direction). HZO domains rotated 180° about the $[111]_{HZO}$ (// $[001]_{STO}$) direction give identical plane-view diffraction patterns ({220} spots with $d_{220}=1.79$ Å in Fig. 3a). Thus, the SAED results are consistent with the existence of four domains rotated by 90° from each other around $[111]_{HZO}$ shown in the pole figure in Fig. 2a. Furthermore, an epitaxial relation between the substrate and the film with $[1-10]_{HZO}//[1-10]_{STO}$, and $[11-2]_{HZO}//[110]_{STO}$, can be clearly identified.

Next, we performed cross-sectional high-angle annular dark-field scanning TEM (HAADF-STEM) analyses on 4 nm and 9 nm thick films, along zone-axes (ZA) defined by $\varphi=0°$ ($[100]_{STO}$), $\varphi=15°$ and $\varphi=45°$ ($[110]_{STO}$). Fig. 3b displays a HAADF-STEM image from a

9 nm thick sample (ZA, $\varphi=45°$), clearly showing the coexistence of majority and minority HZO domains with the $[111]_{HZO}$ and $[001]_{HZO}$ out-of-plane, respectively, in agreement with the x-ray diffraction data. From the fast Fourier transform (FFT) (inset, Fig 3b) across many images, we deduce that $d_{111}$=2.95-3.01 Å and $d_{11-1}$=2.92-2.96 Å. These values ($d_{111} \neq d_{11-1}$) provide a clear confirmation of the non-orthorhombic nature of this phase (see Supplementary Fig. S1 for analysis based on forbidden spots) and support the rhombohedral phase revealed by the synchrotron XRD measurements (Fig. 2b). Fig. 3c displays a HAADF-STEM image from a 4 nm thick sample (ZA, $\varphi=15°$), where we observe the coexistence of different (111)-oriented domains. Notably, (001)-oriented areas are only rarely found at these low thicknesses. (See Supplementary Fig. S2).

We analyzed the chemistry and structure of the interface between HZO and LSMO through energy dispersive spectroscopy (EDS) performed in conjunction with HAADF-STEM. Comparison of the EDS chemical maps with the HAADF-STEM image (ZA, $\varphi=0°$), reveals the presence, at the interface with LSMO, of 2-3 monolayers of HZO that are in a different phase in comparison with the rest of the HZO film (Supplementary Fig. S3). This interfacial HZO phase is completely strained to the substrate (a=3.91 Å), which corresponds to a huge (~8%) in-plane tensile strained t-phase (a=3.60 Å in unstrained t-phase), as shown in Fig. 3d (see also Supplementary Fig. S3c for HAADF simulations). This in-plane tensile strain results in a much lower out-of-plane parameter measured as c/2=2.31-2.44 Å (across several images), compared to the unstrained t-phase (c=5.12 Å). An interfacial t-phase has previously been observed in ALD-synthesized doped-HfO$_2$ samples with TiN electrodes[44], and also on epitaxially strained Y-doped ZrO$_2$ (YSZ) films grown on STO[45]. After the above-mentioned two atomic layers at the interface, the r-phase grows under compressive strain. With increasing thickness, the m-phase in the [001]-orientation appears (Supplementary Fig. S4), also in agreement with the x-ray diffraction data.

In order to test the FE behavior of the films, we obtained polarization versus voltage (P-V) loops through PUND (positive up negative down) measurements (Supplementary Fig. S5).[46] Bi-stable switching and hysteresis loops can be observed in Fig. 4 for 5 nm and 9 nm-thick films. After subtracting the non-switching dielectric contribution, the blue lines only show the FE switching current with coercive field, $E_c$, around 5 MV/cm and 3 MV/cm, respectively, consistent with the expected dependence of $E_c$ with thickness ($d$) as $E_c \sim d^{-2/3}$. The coercive fields are larger than the 1 MV/cm reported in ALD-grown films[32]. This could be due to the dead layers at the interface with the LSMO electrodes as discussed previously.

According to our knowledge, the highest remanent polarization ($P_r$) reported in HZO is 26 μC/cm² (with switching polarization of 45 μC/cm²), using a TiN capping layer of at least 90 nm.[32] In our 5 nm-thick HZO films, $P_r$ also reaches record values of around 34 μC/cm² compared to other epitaxial PLD films[24]. These values are large even when compared with very good conventional perovskites, such as unstrained BaTiO$_3$[47]. However, for the thicker 9 nm film, $P_r$ drops significantly to 18 μC/cm². This finding can be rationalized by our TEM observations showing that the minority (001)-oriented domains (Fig. 3b) and m-phase (non-FE), which are absent or very rarely found in the thinnest sample, gradually appear with increasing thickness (Supplementary Fig. S4). It is interesting to note that the FE loops to be obtained readily after growth, without the technologically inconvenient wake-up cycling required for other HZO systems[48,49] indicating that the FE rhombohedral phase is stabilized by the compressive epitaxial strain.

Finally, we resorted to first-principles simulations to try to identify the rhombohedral HZO phase observed in our films. Since we are not aware of any previous report on a polar rhombohedral polymorph of HZO, we ran a blind search for (meta)stable structures, using the genetic-algorithm approach implemented in USPEX[50–52]. For this we employed standard first-principles methods[53–55], based on density functional theory, and a simulation supercell of 12 atoms (see Methods for details). We ran our search for pure $HfO_2$ and pure $ZrO_2$ compositions in bulk-like conditions. We obtained in both cases an *R3m* structure with a small polarization of the order of 0.1 $\mu C/cm^2$. Furthermore, for the $HfO_2$ composition we also found a second polymorph with *R3* symmetry and a polarization of 41 $\mu C/cm^2$. These rhombohedral phases lie above the *P2₁/c* bulk ground state that is usually discussed in the literature on $HfO_2$, which explains why they have not been previously reported or observed. More precisely, from our calculations we obtain E(*R3m*)-E(*P2₁/c*) = 158 meV/f.u. and E(*R3*)-E(*P2₁/c*) = 195 meV/f.u. for $HfO_2$. Note that the polar *Pca2₁* phase of hafnia discussed in the literature is also more stable than these rhombohedral polymorphs, as we obtain E(*Pca2₁*)- E(*P2₁/c*) = 64 meV/f.u. (in agreement with Ref. 33, which reports 62 meV/f.u. for this energy gap). More details on these structures are provided as Supplementary Information.

We then considered these two rhombohedral structures with the HZO composition, to discover that, upon substitution of Hf by Zr, the *R3* phase loses its stability in favor of the weakly-polar *R3m* polymorph. We studied several Hf-Zr arrangements compatible with the 3-fold symmetry, and observed that the *R3* destabilization occurs in all cases. The *R3m* phase of bulk-like HZO is characterized by $d_{111}$ ~ 2.94 Å and P ~ 1 $\mu C/cm^2$. Then, we examined the effect of epitaxial compression by running simulations for a number of fixed values of the lattice constants in the (111) plane of the HZO *R3m* structure, allowing for the relaxation of atoms and the out-of-plane lattice vector. As shown in Fig. 5, for an epitaxial compression corresponding to an out-of-plane $d_{111}$ ~ 3.25 Å, we observe a clear structural transition to a phase that retains the *R3m* symmetry but is strongly polar, with P ~ 15 $\mu C/cm^2$. Note that this $d_{111}$ value is within the range of what we observe in our thinnest HZO films (for our films with thickness between 1.5 nm and 9 nm, we estimate $d_{111}$ values ranging between ~ 3.27 Å and 2.98 Å, respectively, from the XRD measurements in Fig. 1b). A definite comparison between theory and experiment is not yet possible because of possible phase coexistence in our films as they become thicker. Nevertheless, although our computational models do not include effects (crystallite size; surfaces and interfaces; local deviations from the average composition) that could help to stabilize particular phases, our results do suggest that our predicted *R3m* phase under epitaxial compression, and even the *R3* polymorph predicted in the Hf-rich limit (see Supplementary Fig. S6), may be approximate representations of the rhombohedral structure in our actual samples.

It is important to note that, in general, we can expect epitaxial strain to lower the symmetry of the films; thus, obtaining a rhombohedral unit cell under isotropic epitaxial strain would imply an initially cubic or rhombohedral crystallite. Therefore, the present results allow us to propose a model for the formation of the as-grown FE phase reported here, as follows. PLD growth of the thin films at high temperature enables the in-situ crystallization of HZO. As previously proposed for hafnia-based ferroelectrics, the small particle sizes induce the formation of low-volume fluorite-like phases (either tetragonal or cubic).[38,41,42] A plane-view bright field TEM image (supplementary Fig. S7) from a 9 nm HZO film shows an average grain size of ~10 nm. Our experiments strongly suggest that in the initial stages of the growth, after the formation of a fully coherent, atomically thin, interfacial layer, the internal pressure due to the small particle size favors the undistorted cubic phase. As established by Navrostky et al.[56], the (111) crystal face of $ZrO_2$ and $HfO_2$ is energetically favorable, so cubic crystallites growing with that orientation are expected. Due to the favorable epitaxial relationships induced by the STO/LSMO stack (see Fig. 3a), the growing crystallites are subjected to a large epitaxial compressive strain that elongates the cubic unit cell along the out-of-plane [111] direction, inducing rhombohedral symmetry with a polar unit cell (as shown by the synchrotron XRD and FE characterization).

Increasing the thickness allows the crystal size to grow, relieving the internal pressure,

thus favoring the monoclinic bulk structure. At the same time, the presence of the secondary m-phase also helps to release the elastic energy of the compressively strained structure. Even though XRD only shows traces of the m-phase for thicknesses above ~10 nm, the TEM analysis shows that monoclinic (001)-oriented crystallites are already present in the 9 nm thick film (Supplementary Fig. S4).

To conclude, we show that strain engineering can be used in very thin films of HZO to induce a previously unreported ferroelectric rhombohedral phase, with a large $P_r$ of up to 34 $\mu C/cm^2$. The insight gained in this work provides the missing clues in the understanding of robust ferroelectricity in thin hafnia-based systems and also helps to overcome one of the main issues for their device utilization: the so-called wake-up cycling. Our theoretical calculations predict an even larger polarization for the rhombohedral phases of Hf-rich compositions, and comparable values for epitaxially compressed HZO structures. In addition, this work suggests a pathway to generate large FE polarization in nano-crystallites of simple oxides, whose rich phase diagrams include low volume cubic, tetragonal and rhombohedral phases, and in particular in materials with a clear preference for one specific crystal orientation. These highly-oriented cubic phases can be stabilized during growth and deformed into a polar structure via epitaxial strain. Furthermore, the highly epitaxial growth of ultra-thin FE hafnia-based films on LSMO has great potential for multiferroic tunnel junctions.

**Methods**

**Thin-film synthesis:** Thin films of $Hf_{0.5}Zr_{0.5}O_2$ (HZO) with thicknesses in the range of 1.5-27 nm were grown by pulsed laser deposition (PLD) on $La_{0.7}Sr_{0.3}MnO_3$(LSMO)-buffered (001)-$SrTiO_3$ substrates. A KrF excimer laser with wavelength 248 nm was used to ablate polycrystalline targets of LSMO (purchased by PI-KEM, as bottom electrode) and sequentially, $Hf_{0.5}Zr_{0.5}O_2$ (home made). LSMO was deposited using a laser fluence of 1 J/cm$^2$ and a laser frequency of 1 Hz under a 0.15 mbar oxygen atmosphere and a substrate temperature of 775 °C. A ceramic HZO target (monoclinic $P2_1/c$ phase) was synthesized at 1400 °C by solid state reaction, starting from $HfO_2$ (99% purity) and $ZrO_2$ (99.5% purity) powders. A fluence of 1.1 J/cm$^2$ and repetition rate of 2 Hz was employed to grow the HZO films. The deposition was performed in an oxygen pressure of 0.1 mbar while keeping the substrates at a temperature of 800 °C. After deposition, the film was cooled down to room temperature at a rate of 5 °C/min under an oxygen pressure of 300 mbar.

**X-ray structural charactetrization**: The structure and orientation of the films was characterized by X-ray diffraction (XRD), using a Panalytical X'pert Pro diffractometer operating in two modes: 1) using the line focus of the incident beam for high-resolution *θ-2θ* specular scans, which provide the lattice parameters; 2) using the point focus of the incident beam, for high-intensity/medium-resolution measurements of pole figures, which involve scanning *φ* (azimuthal angle) and *χ* (tilt of the sample plane around the incident beam direction), while the detector is fixed at a particular Bragg reflection. Synchrotron diffraction measurements are performed at the P08 High Resolution Diffraction Beamline in PETRA III, with a wavelength of $\lambda$=1.378 Å, using a Kohzu 6-circle diffractometer and a two-dimensional Pilatus100k detector

**Electron microscopy and energy dispersive spectroscopy:** Cross-sectional and plan-view specimens for electron microscopy were prepared via the standard procedure of mechanical grinding and polishing the samples down to 100 μm, dimpling (Gatan) followed by ion milling (PIPS II) to electron transparency. Electron microcopy and energy dispersive spectroscopy (EDS) were performed on a Titan G2, Cs corrected TEM equipped with an X-FEG and a Super X EDS system with four solid state detectors placed symmetrically along the optical axes. Atomic resolution imaging was performed in the High Angle Annular Dark Field STEM (HAADF-STEM) mode- where the intensity in the image ~ $Z^2$, thus yielding a clear atomic number contrast. Images were analyzed using Image J, TIA-ES Vision and Digital micrograph. EDS was performed simultaneously with image acquisition in the HAADF-STEM mode. Chemical maps were acquired for 30 minutes with conditions optimized and collected at thousands of x-ray photons per second, and analyzed using Bruker e-spirit software.

**Electrical measurements:** LSMO top electrodes were deposited by PLD with the same growth condition as the bottom LSMO electrodes, as described in the synthesis section. LSMO top electrode pads with different sizes were processed by physical etching (ion milling). Planar LSMO/HZO/LSMO capacitors were measured using a ferroelectric tester (AiXACCT, TF analyzer 2000). The ferroelectric response of the films was tested via PUND (positive up negative down) measurements, which are able to separate switching currents from other contributions.

**Simulation methods:** For our structural predictions, we used the USPEX code[50,52,57–59], which is based on genetic algorithms. We applied this method to search for possible rhombohedral phases of $HfO_2$ and $ZrO_2$ in bulk-like conditions. We considered 12-atom cells for $HfO_2$ and $ZrO_2$. After the blind structural search, we ran additional first-principles simulations for $HfO_2$, $ZrO_2$ as well as $Hf_{0.5}Zr_{0.5}O_2$ compositions focusing on the discovered *R3* and *R3m* phases. For the $Hf_{0.5}Zr_{0.5}O_2$ composition, we considered several representative Hf/Zr arrangements that preserve the three-fold rotation axis, using a 72-atom cell.

Our first-principles were carried out using the VASP code[53,55]. We used the PBEsol approximation to density functional theory[54]. Atomic cores were treated within the projector-augmented wave (PAW) approach[60], and we solved explicitly for the following electrons: 5p, 6s, 5d for Hf; 4s,4p,5s and 5d electrons for Zr; 2s and 2p electrons for O. We represented the electronic wave functions in a plane-wave basis, cut-off at 500 eV. For the ground state of the bulk $HfO_2$ ($P2_1/c$) with lattice parameters (a, b, c) of about (5.1, 5.2, 5.2) Å, a mesh of 8x8x8 k-points was used for Brillouin zone integrations; similarly-dense grids were used in all other simulations in this work. When doing the structural relaxation, the calculations were stopped for residual forces below 0.005 eV/A. To simulate the effect of epitaxial strain, we used a hexagonal-like cell with in-plane lattice vectors forming an angle of 120º, and the third lattice vector perpendicular to the plane. The in-plane vectors were fixed at a series of values, smaller ones corresponding to stronger in-plane compression; then, the out-of-plane vector, as well as the atomic positions, were relaxed.

We made extensive use of crystallographic servers[61–63] and the VESTA[64] visualization program. We computed polarizations by comparing the polar structures with properly symmetrized, paraelectric states that we use as reference. Initially we used the Berry phase formalism to compute the polarization from first principles[65], and acquired essentially identical results to those obtained by computing the total dipole based on nominal ionic charges; hence, we used the latter method for most of our later calculations.

**Acknowledgements:**

We are grateful to Sergey Volkov and Florian Bertram for their help at the P08 beamline in Petra III (DESY0-Hamburg). YW and BN are grateful for China Scholarship Council and Van Gogh travel grant. PN and BD would like to acknowledge a public grant overseen by the French National Research Agency (ANR) as a part of the "Inverstissements d'Avenir program (Grant No. ANR-10-LABX-0035, Labex Nanoscalay). HJZ and JÍ acknowledge the support of the Luxembourg National Research Fund through the PEARL (Grant FNR/P12/4853155/Kreisel COFERMAT) and CORE (Grant FNR/C15/MS/10458889 NEWALLS) programs.

Figures:

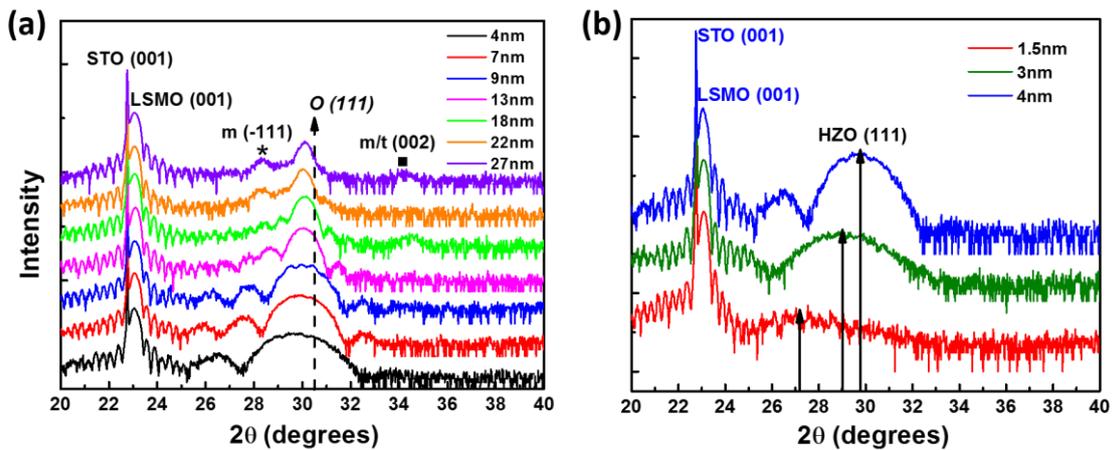

**Figure 1: XRD structural characterization of Hf$_{0.5}$Zr$_{0.5}$O$_2$ (HZO) films on LaSrMnO$_3$-buffered 001-oriented SrTiO$_3$ (001-STO/LSMO). a,** Specular X-ray diffraction pattern of HZO films with thicknesses ranging from 4 nm to 27 nm. **b,** Specular X-ray diffraction patterns of HZO films with thicknesses ranging from 1.5 nm to 4 nm.

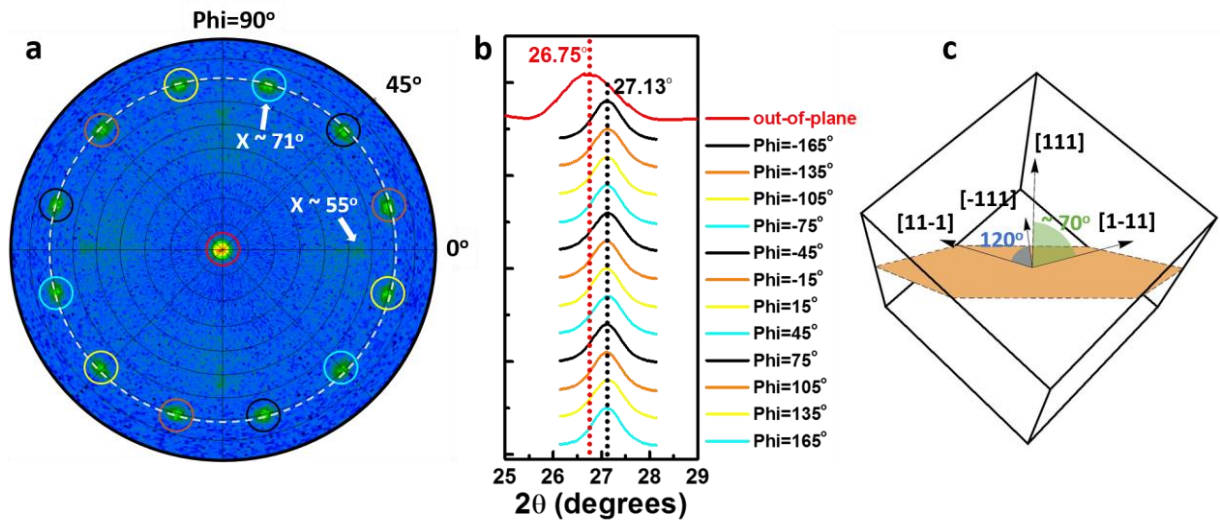

**Figure 2: Domain configuration and symmetry. a,** Pole figure around the (111) peak of a 9 nm HZO film at 2θ=29.98°. The radial direction represents χ, which ranges between 0° and 90°, while the azimuthal direction represents φ, with a (0° - 360°) range. **b,** 2theta scans of the 13 peaks in the pole figure measured at the P08 High-Resolution Diffraction Beamline in PETRA III (DESY) with a wavelength of λ=1.378 Å. **c,** Sketch of the proposed rhombohedral structure of HZO film with polarization along the [111] direction.

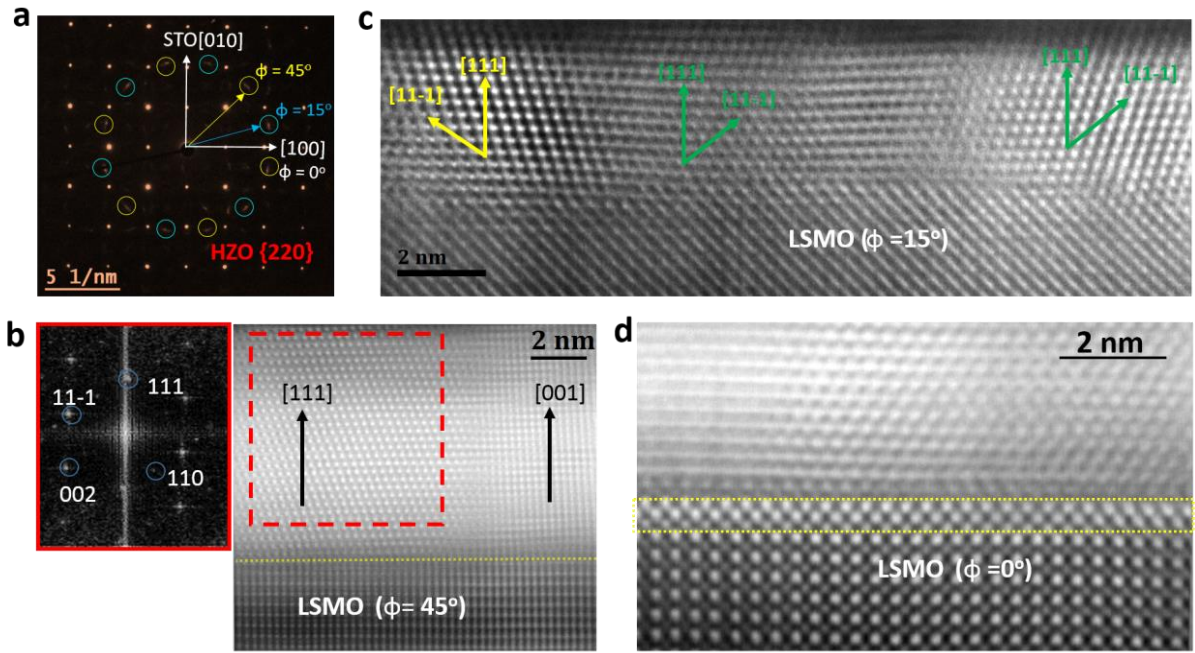

**Figure 3: Electron microscopy characterization. a,** Plane view selected area electron diffraction (SAED) pattern from a 9 nm thick HZO sample. The out of plane direction is [111] (zone-axis). The {220} spots ($d_{220}$=1.79 Å), corresponding to at least two different domains (yellow and blue circles, respectively) rotated by 90º with respect to each other, can be clearly identified. **b,** Cross-sectional (x) HAADF-STEM image (corrected by sample drift) of the 9 nm thick HZO film, observed along the [110] zone of the substrate (φ =45º); (inset, left) Fourier transform of the [111] domain; **c,** Representative x-HAADF STEM image (drift-corrected) of a 4 nm thick film, observed along the zone axis defined by φ =15º. (d) HAADF-STEM image observed along STO [100], revealing a clear interfacial t-phase of HZO (see EDX in Supplementary Fig. S3).

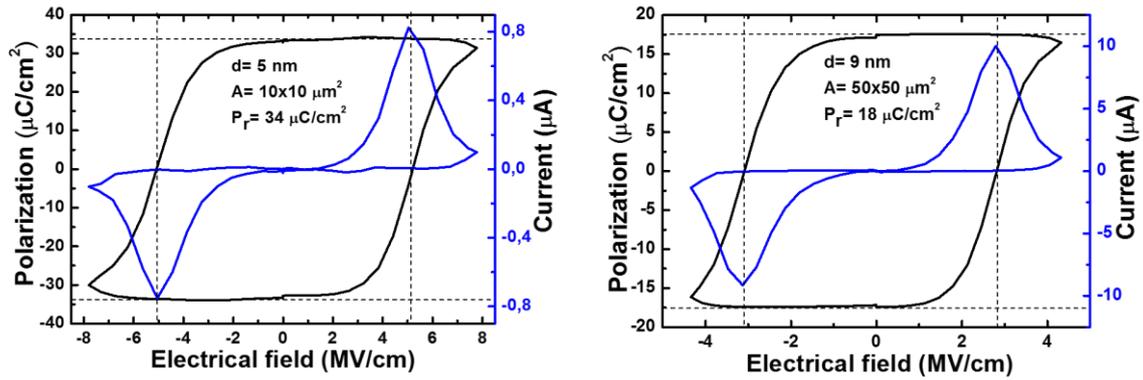

**Figure 4: Ferroelectric characterization.** PUND measurements (see Supplementary information) of 5 nm- and 9 nm-thick films under an electric field of frequency 1 kHz. The blue lines are the current (I) *versus* voltage (V) curves after extracting the dielectric response; the black lines are the integrated signal giving rise to the corresponding polarization (P) *versus* voltage (V) hysteresis loops.

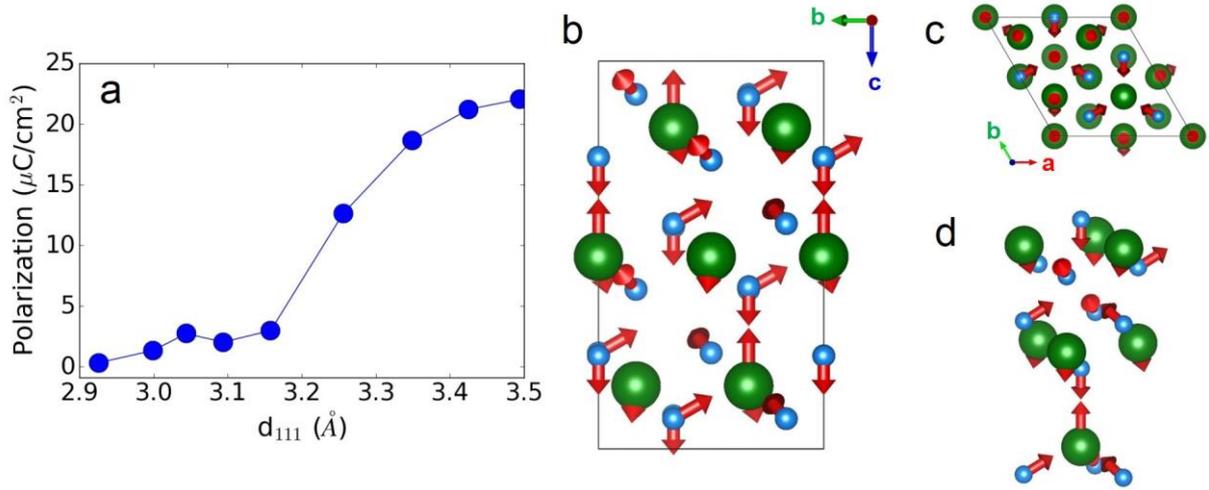

**Figure 5: Theoretical calculations and proposed structure. a,** Computed polarization of the R3m phase of HZO as a function of $d_{111}$. Note that we simulate epitaxial strain, and thus control $d_{111}$, as described in the text, where larger $d_{111}$ values correspond to smaller in-plane lattice constants. **b,c,** Two views of the *R3m* phase obtained for epitaxially-compressed HZO and $HfO_2$. Green (cyan) spheres represent Hf/Zr (O) atoms. The arrows show the polar distortion with respect to a reference paraelectric structure with *R-3m* symmetry, which we deduce by appropriately symmetrizing the *R3m* state. **d,** Detailed view of the Hf/Zr—O groups that characterize the *R3m* structure, allowing the main polar displacements, and how they preserve the 3-fold axis, to be more easily appreciated.

# Supplementary information

# A rhombohedral ferroelectric phase in epitaxially-strained Hf$_{0.5}$Zr$_{0.5}$O$_2$ thin films

Yingfen Wei[1], Pavan Nukala[2], Mart Salverda[1], Sylvia Matzen[3], Hong Jian Zhao[4], Jamo Momand[1], Arnoud Everhardt[1], Graeme R. Blake[1], Philippe Lecoeur[3], Bart J. Kooi[1], Jorge Íñiguez[4], Brahim Dkhil[2], Beatriz Noheda[1],*

[1]Zernike Institute for Advanced Materials, University of Groningen, 9747 AG Groningen, The Netherlands

[2]Laboratoire Structures, Propriétés et Modélisation des Solides, CentraleSupélec, CNRS-UMR8580, Université Paris-Saclay, 92295 Châtenay-Malabry, France.

[3]Center for Nanoscience and Nanotechnology, CNRS-UMR 9001, Université Paris-Saclay, 91405 Orsay, France

[4] Materials Research and Technology Department, Luxembourg Institute of Science and Technology (LIST), 5 avenue des Hauts-Fourneaux, L-4362 Esch/Alzette, Luxembourg

* E-mail: b.noheda@rug.nl;

**1. SAED simulations of the polar orthorhombic phase (Pca2$_1$):** SAED simulations of the orthorhombic phase commonly reported in literature were obtained through multislice methods using JEMS software. From these simulations (Fig. S1a), we note that in the <110> zone, one of the {001} or {1-10} spots are forbidden. However, in Fig S1b, when compared with the fast-Fourier transform of the cross-sectional HAADF-STEM image of a [001] domain, we see that both the {001} and the {1-10} spots are present, reinforcing the conclusion that our films, which are polar, exhibit a different symmetry to the commonly reported polar o-phase.

**2. Interfacial tetragonal phase**: Using a combination of EDS and HAADF-STEM (Fig. S3, HAADF-simulation in Fig. S3c), we deduced that the first couple of monolayers belong to a tensile strained tetragonal phase.

**3. Evolution to a monoclinic (bulk) phase with thickness**: HAADF image simulations were performed on all the phases (P2$_1$/c, rhombohedral phases obtained from our theoretical results, Pca2$_1$). On the 9 nm thick sample, using these simulations, we could clearly deduce the evolution from an r-phase to a bulk m-phase with increasing thickness (Fig. S4). Thus, our TEM results indicate the existence of non-polar monoclinic crystals in the 9 nm thick samples, rationalizing our observations of lower polarization values as compared to thinner films .

**4. PUND measurement (positive up negative down):** As shown in Fig. S5, the first pulse is the pre-write pulse. After the preset pulse, the first read pulse is a positive switching pulse, the second is an unswitched pulse, the third is a negative switched pulse, and the last is a negative unswitched pulse. I$_{realFE}$ (the current from the real ferroelectricity switching) = I$_s$ (the current from the switching pulse) - I$_{non}$ (the current from the unswitched pulse). The current (blue) in Fig. 4 combines positive (I$_1$-I$_2$) and negative (I$_3$-I$_4$) parts and is plotted as a function of electric field. Polarization (P) is calculated from the formula $\int I dt/A$ (I :current; t: time; A: electrode area).

**5. Additional details on the predicted rhombohedral phases:** In the table below we detail the crystallographic structures that were obtained from our first-principles simulations and are representative of the polar rhombohedral phases discussed in this work. We give the structures obtained from HfO$_2$ simulations; obtaining the corresponding structures for the HZO composition is trivial, by performing the corresponding substitution of atoms and completing a further structural relaxation. Finally, Fig. S6 shows two views of the *R3m* and *R3* phases of HfO$_2$.

| Bulk HfO$_2$, *R3* (No. 146) |
| --- |
| a=b=7.106 A; c=9.016 A; α=90º; β=90º; γ=120º |
| Hf  9b  0.86366  0.67270  0.24007 |
| Hf  3a  0.00000  0.00000  0.55493 |
| O   9b  0.13334  0.32314  0.14948 |
| O   9b  0.39917  0.24614  0.01541 |
| O   3a  0.00000  0.00000  -0.08408 |
| O   3a  0.00000  0.00000  0.32548 |
| Bulk HfO$_2$, *R3m* (No. 160) |
| a=b=7.134 A; c=8.741 A; α=90º; β=90º; γ=120º |
| Hf  9b  0.83335  0.16665  0.25089 |
| Hf  3a  0.00000  0.00000  0.58415 |
| O   9b  0.14966  0.85034  0.15904 |
| O   9b  0.48699  0.51301  0.32788 |
| O   3a  0.00000  0.00000  0.85998 |
| O   3a  0.00000  0.00000  0.35364 |
| HfO$_2$, compressed in-plane, *R3m* (No. 160) |
| a=b=6.683 A; c=10.041 A; α=90º; β=90º; γ=120º |
| Hf  9b  0.82944  0.17056  0.27065 |
| Hf  3a  0.00000  0.00000  0.55278 |
| O   9b  0.14881  0.85119  0.15456 |
| O   9b  0.48218  0.51782  0.32402 |
| O   3a  0.00000  0.00000  0.86126 |
| O   3a  0.00000  0.00000  0.34948 |

**6. Particle size estimation:** Particle size statistics on the 9 nm thick HZO film were estimated from a bright-field TEM image obtained in plane view (Fig. S7a). We utilized the particle analysis procedure prescribed by digital micrograph to obtain a median particle area of 84 nm$^2$ (or diameter of ~10 nm), similar to the film thickness. Similarly, in a 4 nm film, we found grain sizes of 3-4 nm (Fig. 3c in the main manuscript).

**Supplementary figures**

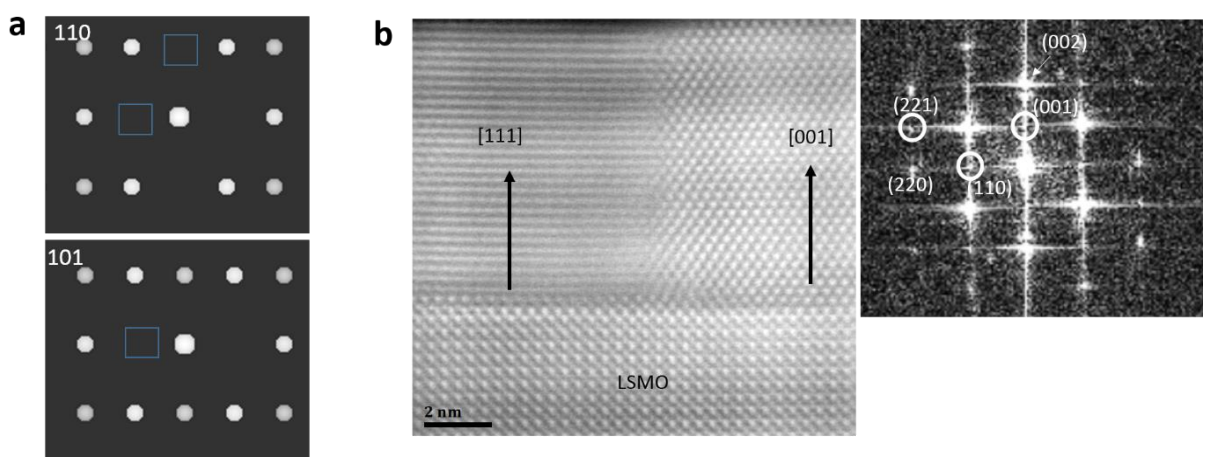

**Figure S1. SAED simulations of the polar orthorhombic phase (Pca2$_1$). a,** SAED diffraction simulations of Pca2$_1$ phase, along the [110] and [101] zone axes. Along the [110] zone, we note that both the (001) and the (1-10) spots are forbidden. Along the [101] zone, (10-1) is forbidden, and along [011] (not shown), the (100) spots are forbidden. **b,** x-HAADF-STEM image from a 9 nm thick sample in the with zone axis: [100]$_{STO}$. Domain coexistence (majority c-axis= [111], and minority c-axis= [001] can be clearly observed. Inset on the right shows FFT in which some spots from the 001 domain are indexed and circled. In an orthorhombic Pca2$_1$ phase, one of the (001) or the (110) spots should be forbidden (by symmetry). However, we see both of them, clearly suggesting that we have a different ferroelectric phase than the commonly reported Pca2$_1$ phase.

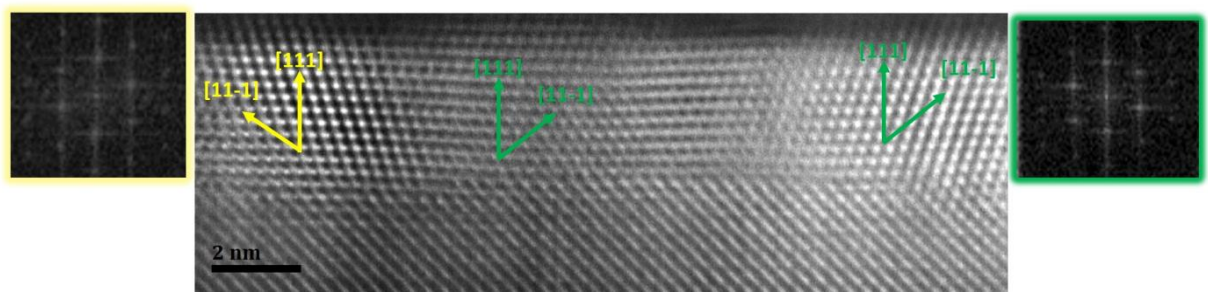

**Figure S2. Domains in 4nm thick sample.** Representative HAADF-STEM image from a 4 nm thick sample (Fig. 3c of the manuscript), with zone axis defined by $\varphi=15°$. At least two domains (green and yellow, 180° rotated), with out of plane [111] can be seen. The [001] out of plane domains in this sample were very rarely found (as opposed to the 9 nm thick sample). Insets on the left and right show FFTs of the yellow and green domains respectively. $d_{111}$~2.95-3.01 Å, $d_{11\text{-}1}$~2.93-2.97 Å in both the domains.

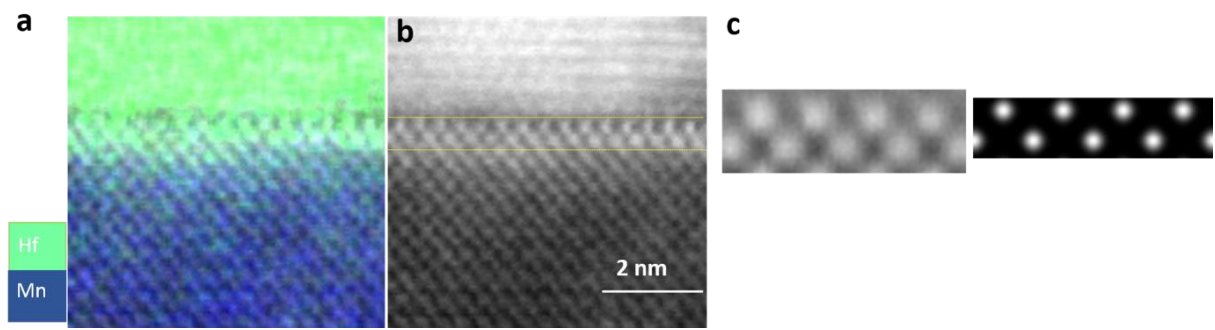

**Figure S3 : Interfacial tetragonal phase. a,** Hf and Mn chemical maps obtained via energy dispersive spectroscopy in a 9 nm thick sample. **b,** Corresponding HAADF-STEM ($\varphi=0°$) image. Comparing (a) with (b) clearly shows that interfacial phase is indeed chemically HZO, and fully strained with respect to the substrate, and in a different phase compared to the rest of the HZO film. It can be matched to an ~8% in-plane tensile strained t-phase of HZO. **c,** HAADF simulations.

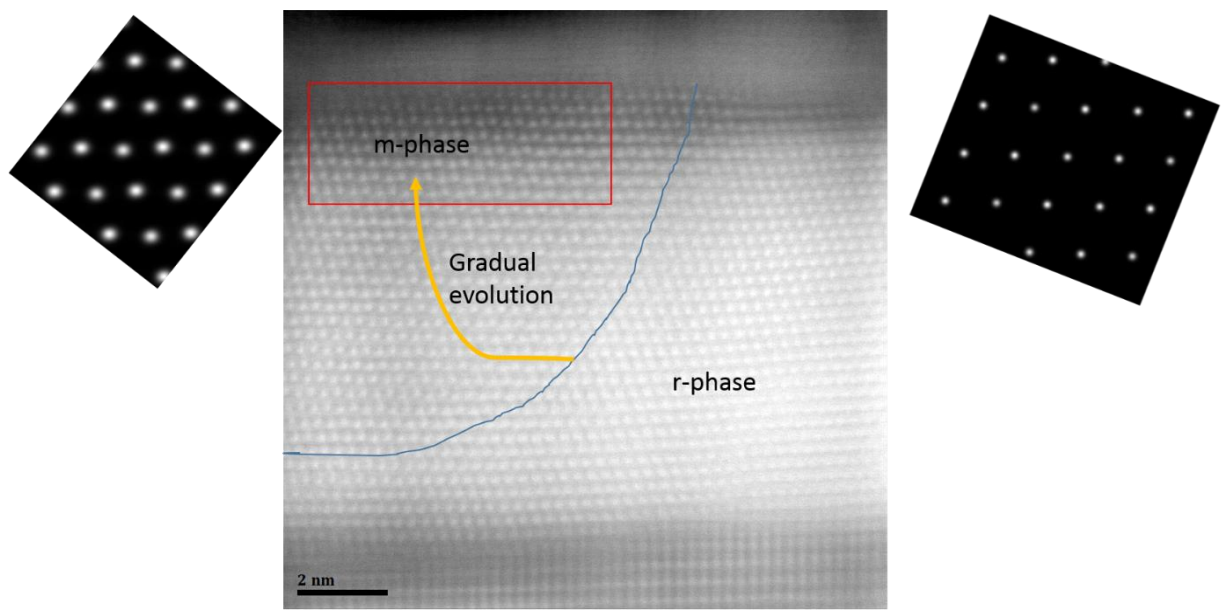

**Figure S4**: **Evolution to a monoclinic (bulk) phase with thickness.** HAADF-STEM cross-section image (ZA, φ=45°) of a 9 nm HZO sample showing a gradual evolution of r-phase into bulk-m phase with increasing thickness. (inset left), HAADF-STEM simulation of an m-phase clearly showing the zig-zag atomic positions along [112] (horizontal direction), (right) Corresponding HAADF-STEM image simulation. Renditions of various rhombohedral structures obtained from our structural search are shown in Fig. S6.

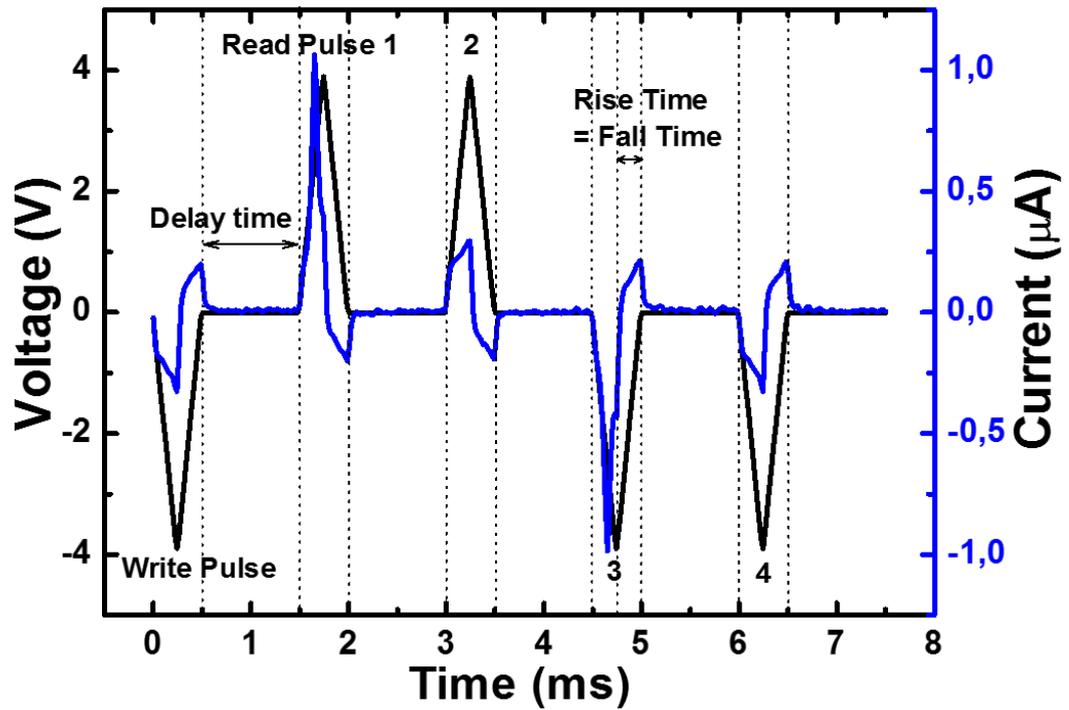

**Figure S5: PUND (Positive up negative down) measurements.** PUND (positive up negative down) measurement on a 5 nm thick film with 1 kHz electric field frequency.

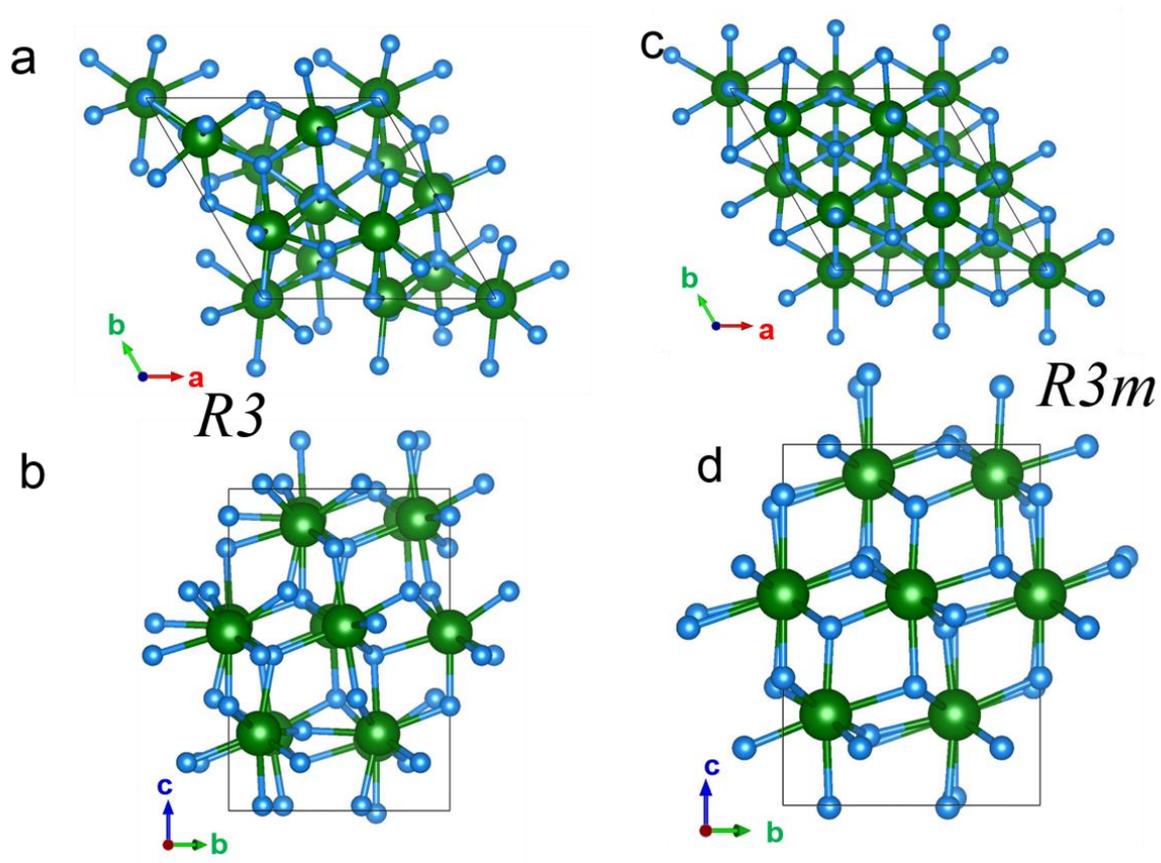

**Figure S6: Rhombohedral polar phases of bulk HfO$_2$ obtained in our structural search. a,b,** show two views of the *R3* phase, while **c,d** show two views of the *R3m* phase. Hexagonal axes are shown as reference. Hf and O atoms are represented by green and blue spheres, respectively.

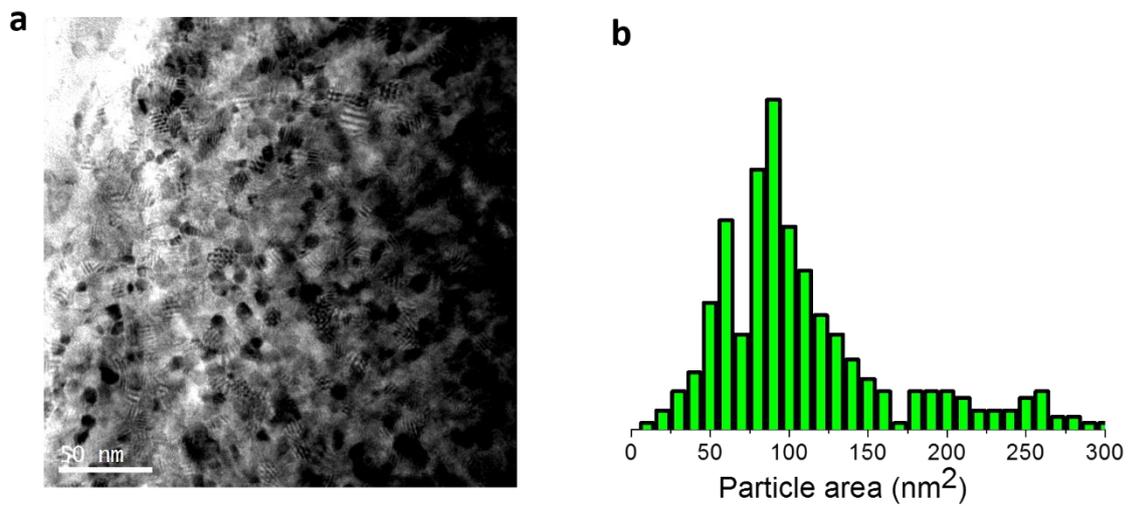

**Figure S7. Particle size estimation. a,** Plane view bright-field TEM image of a 9 nm HZO sample. **b,** Grain size distribution from the image in (a) revealing a median grain size of ~10 nm (median grain area of 84 nm$^2$).